# A Two-Stage Framework for Fast Proton Spot Map Generation in Pencil Beam Scanning Prostate SBRT Planning

*Xueyan Tang, PhD, Hok Wan Chan Tseung, PhD, Mark Pepin, PhD, Jiasen Ma, PhD, David M. Routman, MD, Doug J. Moseley, PhD, Brandon Reber, PhD, Jed E. Johnson, PhD, Jing Qian, PhD*

Department of Radiation Oncology, Mayo Clinic, Rochester, MN

## Abstract

**Background:** In pencil beam scanning (PBS) proton therapy, treatment plans are encoded as proton spot maps (PSMs). Although deep-learning methods can rapidly predict three-dimensional (3D) dose distributions from patient anatomy and planning intent, there is still no widely adopted direct, non-iterative approach to convert doses into physically deliverable spot patterns. This limits end-to-end automation in treatment planning and adaptive replanning for PBS.

**Purpose:** We developed and retrospectively evaluated GenSpot, a two-stage framework that infers a deliverable PSM from a given CT and dose distribution in a single-institution cohort of prostate stereotactic body radiation therapy (SBRT) plans. Our primary goal was to determine whether GenSpot could generate PSMs whose Monte Carlo (MC) dose distributions closely match those calculated from the clinically delivered PSMs.

**Methods:** We introduced a physics-informed projected proton spot map (PrPSM) representation that projects individual proton spots through the CT using water-equivalent thickness and percentage depth dose (PDD) information, aligning the spots with the CT and dose voxel grid while preserving a linear relationship with spot weights. Our dataset comprised 1,036 fields from 259 prostate SBRT plans, split into training, validation, and test sets in an 80%/10%/10% ratio. A 3D SwinUNETR model was trained to predict PrPSMs from CT and dose inputs. Field-specific PSMs were then reconstructed from the predicted PrPSMs using column-wise nonnegative Lasso regression with precomputed PDD curves. MC dose distributions from GenSpot and clinical PSMs were compared using voxel-wise mean absolute error (MAE), 3D gamma analysis (3%/3 mm, 10% low-dose threshold), and dose-volume histogram (DVH) metrics at the composite plan level.

**Results:** Within the held-out test dataset, the selected SwinUNETR model achieved a mean PrPSM MAE of 0.06 ± 0.02 (normalized units) with high structural similarity to projected clinical PrPSMs. At the dose level, MC doses from GenSpot PSMs showed low MAE (0.07 ± 0.03 Gy in the nonzero-dose region) and high gamma passing rates (0.90 at the field level and 0.97 at the plan level). Composite-plan DVH metrics differences were generally within 1 Gy for targets and organs at risk, although the high-dose tail of the clinical target volume showed a modest systematic increase. Spot map complexity, measured by the number of energy layers and spot counts, was similar to that of clinical plans, with slightly more spots for GenSpot PSMs. Mean PrPSM prediction and PSM reconstruction times were approximately 0.02 s and 2.1 s per field, respectively.

**Conclusions:** In this single-institution prostate SBRT cohort, GenSpot reconstructed PSMs that satisfied machine delivery constraints from CT and dose, and the resulting MC dose distributions closely matched those from clinically delivered PSMs. To our knowledge, this is the first practical physics-informed framework to recover deliverable PSMs directly from dose and anatomy, providing a promising dose-to-spots bridge toward more automated PBS workflows. Broader multi-site validation and comparison with established dose-mimicking approaches will be needed before routine clinical use.

## 1 Introduction

Pencil beam scanning (PBS) proton therapy is widely used as a standard delivery technique, providing highly conformal dose distributions and the potential to reduce normal tissue toxicity compared with conventional photon radiotherapy.[1] These advantages arise from the Bragg peak and the steep distal dose fall-off of protons.[2] The same sharp gradients, however, increase sensitivity to setup and range uncertainties, particularly for intensity-modulated proton therapy (IMPT) plans with highly modulated spot weights.[3] As a result, robust optimization and extensive plan verification have become routine in clinical PBS workflows, at the cost of larger margin, greater algorithmic complexity and substantial computational demands in both treatment planning and quality assurance.[4,5]

In current practice, a PBS plan is encoded by a proton spot map (PSM) that specifies the position, nominal energy, and weight of each spot; a single field may contain tens of thousands of spots. The PSM is the direct input to the delivery system and represents the endpoint of inverse planning. Although modern treatment planning systems can generate high quality IMPT plans, the underlying optimization remains iterative, nonconvex, and labor intensive.[6] Achieving institutional dose-volume and robustness criteria often requires repeated cycles of objective tuning, replanning, and manual review, which is especially burdensome in time-sensitive settings such as adaptive proton therapy.[7,8]

Automation and artificial intelligence (AI) have already transformed treatment planning in photon radiotherapy. For intensity-modulated radiation therapy and volumetric modulated arc therapy, knowledge-based planning and machine-learning-driven auto-planning can reproduce expert-level plan quality, reduce inter planner variability, and shorten planning time.[9-11] Similar strategies are increasingly being explored in proton therapy, including knowledge-based IMPT, automated robust multicriteria optimization, and deep reinforcement learning.[12-14] In most current proton implementations, however, AI is used to support or accelerate steps within the conventional inverse-planning workflow rather than to replace optimization altogether. A natural direction for further acceleration is to reduce reliance on iterative optimization by generating deliverable spot maps more directly from an intended dose distribution, followed by rapid dose calculation and limited local refinement.

A complementary line of research reframes planning as a dose prediction problem. Instead of specifying objectives and constraints and solving a large-scale inverse problem, dose prediction models learn a direct mapping from patient anatomy and planning intent to a three-dimensional (3D) dose distribution that resembles the output of an optimized plan. In proton therapy, deep-learning models have been used to estimate PBS dose from CT or synthetic CT and structures, sometimes with only high-level beam descriptors, and can achieve close agreement with clinical IMPT plans in terms of dose-volume histogram (DVH) metrics and gamma analyses.[15,16] In practice, however, these predicted doses are usually used for plan comparison or decision support rather than as directly deliverable plans. Converting a predicted dose into a clinically usable plan still requires an additional inverse step, such as dose-mimicking optimization, which reintroduces an ill-posed and computationally expensive problem.

A method capable of inferring a deliverable PSM directly from dose and anatomy would therefore fill a critical gap in the dose-to-spots pipeline. Such an approach could enable rapid adaptive replanning by reconstructing updated spot maps from adapted dose distributions, or provide high-quality initial spot configurations that reduce the burden of subsequent inverse optimization. The inverse mapping is challenging for several reasons. The relationship between spot weights and dose depends strongly on patient-specific heterogeneities and beam geometry and is therefore nonlinear in practice. The mapping is also fundamentally non-unique, because multiple distinct spot configurations can yield nearly indistinguishable dose distributions in clinically relevant regions. Most importantly, PSMs and medical images inhabit different spaces: spot maps are parameterized by spot position and nominal energy, whereas CT and dose are defined on a 3D voxel grid. This domain mismatch complicates the use of standard image-based neural networks for spot prediction and motivates an intermediate representation that is compatible with voxel-based learning while remaining tightly linked to spot intensities.

Here we present GenSpot, a deep-learning framework that reconstructs PSMs within machine delivery constraints directly from CT and a 3D dose distribution. GenSpot is built around a physics-informed projected proton spot map (PrPSM) representation that embeds beam-specific information into voxel space while preserving a linearity with respect to spot weights.[17] A 3D transformer-based network predicts PrPSM from CT and dose, and a subsequent sparse linear regression step recovers spot energies and weights to form a GenSpot PSM. We retrospectively evaluate the framework in a single-institution cohort of prostate stereotactic body radiation therapy (SBRT) proton plans, comparing Monte Carlo (MC) dose distributions from GenSpot-reconstructed PSMs with those from clinically delivered PSMs. By providing a direct dose-to-spots mapping in this setting, GenSpot is intended as a building block toward more automated and efficient PBS planning and adaptive workflows.

## 2 Materials and Methods

### 2.1 Data Cohort

The study cohort consisted of 259 clinically delivered prostate SBRT plans generated for treatment on a Hitachi Probeat-V Proton Beam Therapy System at our institution. Among these, 203 plans used a single prescription level of 3800 or 4000 cGy delivered in five fractions. The remaining 56 plans used a simultaneous integrated boost (SIB), most commonly 3800 cGy to the primary target volume and 4000 cGy to the boosted volume. Under our institutional prostate planning standard, each plan contained four fields, yielding a total of 1036 individual fields.

For each plan, we collected the noncontrast planning CT scan, the 3D dose distributions, target and organs at risk (OAR) contours, and the corresponding clinically delivered proton spot maps (PSMs). CT scans were acquired on Siemens SOMATOM Definition AS 20 scanners at 120 kVp, a slice thickness of 1 mm, and an iterative metal artifact reduction algorithm for reconstruction. All clinical dose distributions were calculated using the same clinically commissioned graphics processing unit (GPU)-based Monte Carlo (MC) proton transport simulator used for routine proton planning at our center.[18] The same MC configuration was used to recompute dose from both clinical and GenSpot PSMs to ensure a fair comparison.

CT images, dose distributions, and contours were stored in Digital Imaging and Communications in Medicine (DICOM) format. For each field, the PSM was exported as a 3D array, in which the first two dimensions indexed spot coordinates on the isocenter plane and the third dimension indexed nominal energy. Array entries represented the number of protons assigned to each spot. PSMs were stored as text files along with machine delivery parameters, including isocenter location, gantry angle, couch angle, and range-shifter status.

The study was approved by the institutional review board (IRB 23-007140; approved September 5, 2023). All patient data were anonymized prior to analysis, and written informed consent was obtained from all participants.

### 2.2 Data Preparation

#### 2.2.1 Preprocessing

Figure 1 summarizes the preprocessing workflow. CT images, dose distributions, and structure contours were exported in DICOM format and converted to Neuroimaging Informatics Technology

Initiative (NIfTI) format to facilitate downstream processing.[19] Spatial coordinates were encoded in the NIfTI affine matrix, enabling direct mapping between image voxels and patient-space locations.

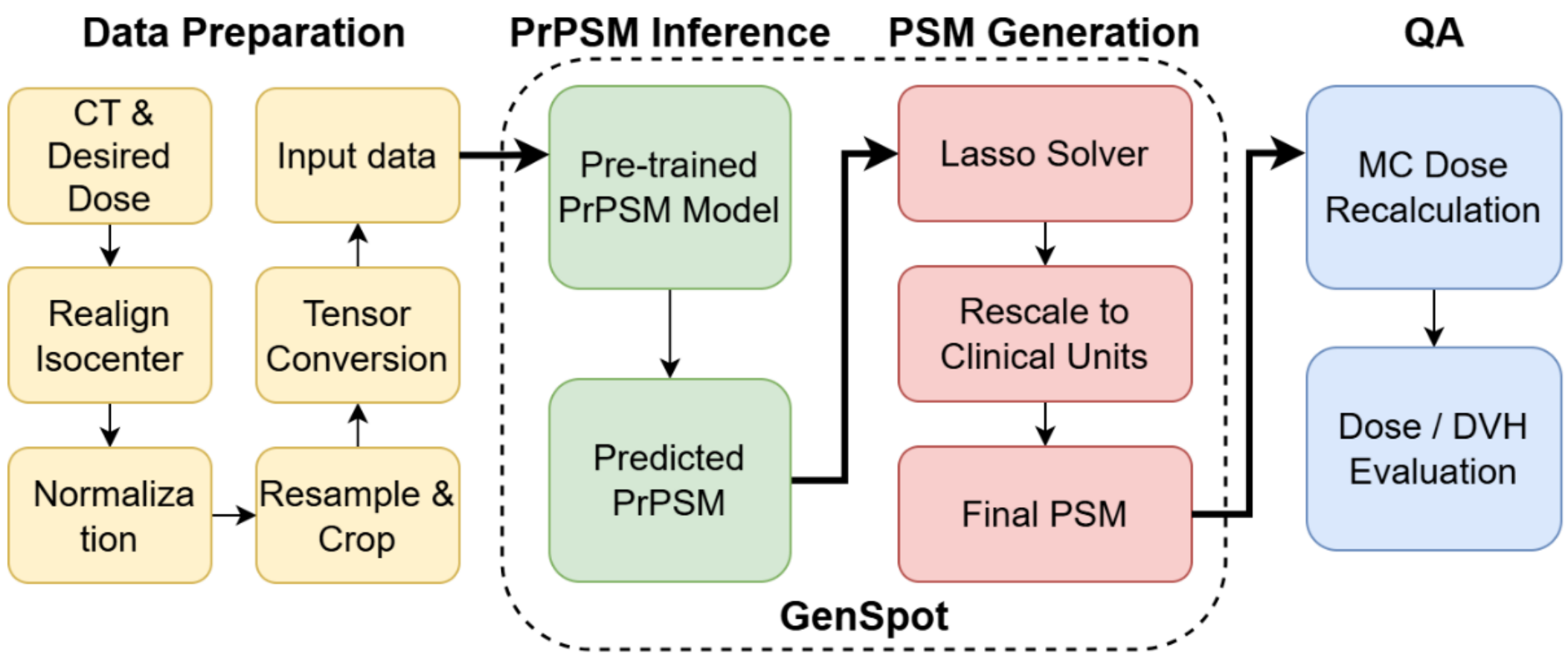

Figure 1. Overview of the GenSpot workflow.

All images were rigidly translated so that the treatment plan isocenter coincided with the image center. Volumes were then resampled to an isotropic spatial resolution of 2.5 mm, intensity normalized, and cropped to a fixed size of 256 × 256 × 256 voxels. This field of view encompassed the prostate, relevant OARs, and beam paths in all cases. Each volume was subsequently rotated about the image center according to the gantry and couch angles to obtain a beam's-eye-view (BEV) representation, ensuring that in-plane coordinates were aligned with the x and y coordinates used in the PSMs.[17]

CT Hounsfield units (HUs) were clipped to [−1000, 1000] and divided by 1000, resulting in normalized values in [−1, 1]. Dose distributions were normalized on a per-field basis to the range [0, 1] by dividing by the maximum field dose. The same scaling factor was applied to the corresponding PSM so that the relative relationship between proton numbers and delivered dose was preserved during training. We chose per-field rather than global normalization to accommodate the substantial variation in beam weights across fields. When MC dose was recalculated from reconstructed PSMs, the original field maximum dose of input dose map was restored to obtain physically meaningful absolute doses. In this study, the input dose to GenSpot was always the MC dose from the clinical PSM, providing an unambiguous absolute scaling factor for each field.

Contours of OARs (rectum and bladder) were converted into binary masks, with voxels inside the contoured structure assigned a value of 1 and all other voxels assigned 0. Target volumes were encoded in a single mask that captured multiple prescription levels. For SIB plans, all clinical target volumes (CTVs) were combined: voxels in the highest dose level were assigned a value of 1, while voxels in lower dose levels were assigned scaled values proportional to their prescription relative to the highest level. These masks were not provided as network inputs; rather, they were used for evaluation and to define the regression region during PSM reconstruction.

### 2.2.2 Construction of Projected Proton Spot Maps (PrPSM)

Rather than training directly on PSMs, we defined a projected proton spot map (PrPSM) representation that is co-registered with the CT and dose volumes. In this formulation, protons are assumed to travel along straight rays in the beam direction, and each spot in the original PSM is projected through the CT volume. The projection incorporates basic beam physics via the water-equivalent thickness (WET) along each ray and the CT-derived stopping power ratio, yielding a simplified surrogate for proton fluence that embeds spot information in patient-specific anatomy. Lateral scattering and other higher-order transport effects are neglected, so the PrPSM is intended as an approximate fluence surrogate rather than a full dose model. By construction, it provides a direct spatial correspondence among the PSM, CT, and dose, making it a convenient intermediate representation for the AI model and reconstruction pipeline.

To construct the PrPSM, spots in the PSM are processed sequentially and accumulated into a 3D volume. Each spot's in-plane coordinates (x, y), defined at the isocenter plane, are mapped to voxel indices using a 2.5 mm in-plane spacing, with the isocenter aligned to the image center (indices (128, 128, 128)). Spot size is neglected and each spot is assigned to a single in-plane voxel. Under the straight-line assumption, the spot is then projected along the beam axis, forming a column of voxels along z direction at fixed (x, y) in the BEV coordinates.

For each voxel along this column, the PrPSM intensity is given by

$$PrPSM(x,y,z) = \sum_{s \in S_{x,y}} n_s \cdot PDD\big(E_s, WET(x,y,z)\big), \tag{1}$$

where $S_{x,y}$ is the set of all spots whose BEV coordinates coincide with (x, y), $n_s$ is the number of protons in spot $s$, and $E_s$ is its nominal energy. The function $PDD\big(E_s, WET(x,y,z)\big)$ is the PDD for energy $E_s$, evaluated at the cumulative WET from beam entrance to that voxel.

The local WET increment is computed as the product of voxel thickness and the stopping power ratio derived from the voxel's CT HU. Cumulative WET at a given depth is obtained by summing these increments from beam entrance to the voxel of interest. When a range shifter is present, its effect is modeled as a uniform WET offset along the beam path. Summing the contributions from all spots yields the final PrPSM volume, which appears as a set of straight, parallel columns with depth-varying intensity along the beam direction. (Figure 2b).

Figure 2 compares representative slices of the dose distribution, the corresponding PrPSM, and the original PSM array. In the dose and PrPSM volumes, the third dimension represents physical depth along the beam path, whereas in the PSM array the third index corresponds to nominal energy. The PrPSM therefore exhibits much clearer spatial and intensity correspondence with the dose than the original PSM representation.

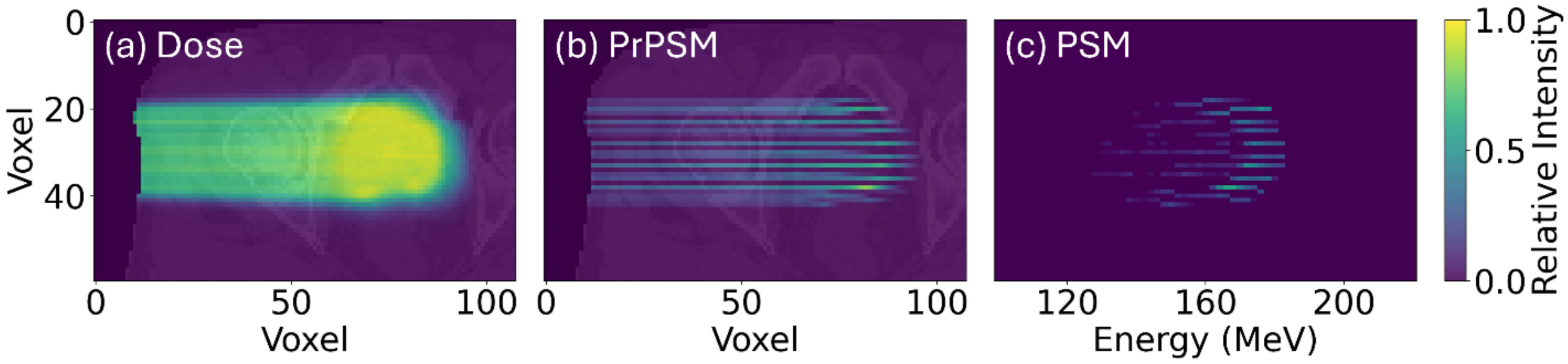

Figure 2. Representative slices of (a) dose, (b) PrPSM, and (c) PSM.

**2.3 AI-based PrPSM Prediction from CT and Dose**

In prior work, we showed that CT anatomy, dose, and PrPSM are strongly coupled by training a deep-learning model to calculate dose from CT and PrPSM. [17] If a network can reliably learn the forward mapping from (CT, PrPSM) to dose, then the inverse pair (CT, dose) should contain sufficient information to infer PrPSM. Motivated by this, we formulated and evaluated the inverse problem of predicting PrPSM directly from CT and dose.

Throughout this manuscript, GenSpot denotes the full pipeline that (i) predicts PrPSM from CT and dose and (ii) reconstructs a PSM from the predicted PrPSM by linear regression (Figure 1). For a given field, the reconstructed spot map produced by this pipeline is referred to as the GenSpot PSM. When transported through the MC simulator, the resulting dose is termed the dose from the GenSpot PSM; the corresponding dose calculated from the clinically delivered PSM is termed the dose from the clinical PSM. In this study, the input dose to GenSpot was always the clinical MC dose, allowing the evaluation to isolate the dose-to-spots mapping without additional uncertainty from an upstream dose-prediction model.

2.3.1 Model Architecture

For PrPSM prediction within GenSpot, we used the SwinUNETR architecture, a transformer-based encoder-decoder previously employed for proton dose calculation (Figure 3). [17,20] The network accepts a two-channel input formed by concatenating the CT and dose volumes and outputs a single-channel PrPSM volume at the original spatial resolution.

SwinUNETR integrates a hierarchical Swin Transformer encoder, which uses window-based and shifted-window self-attention to capture both local structure and long-range dependencies, with a UNet-style decoder that reconstructs voxelwise outputs via progressive upsampling and skip connections. In the encoder, the input volume is partitioned into non-overlapping 3D patches that are embedded into a latent feature space. Subsequent Swin Transformer stages with patch merging progressively build multi-scale feature representations. The decoder mirrors this hierarchy: deconvolution layers progressively upsample the feature maps, which are concatenated with corresponding encoder features through skip connections and further refined by convolutional residual blocks. A final 1 × 1 × 1 convolution layer maps the aggregated features to the predicted PrPSM volume.

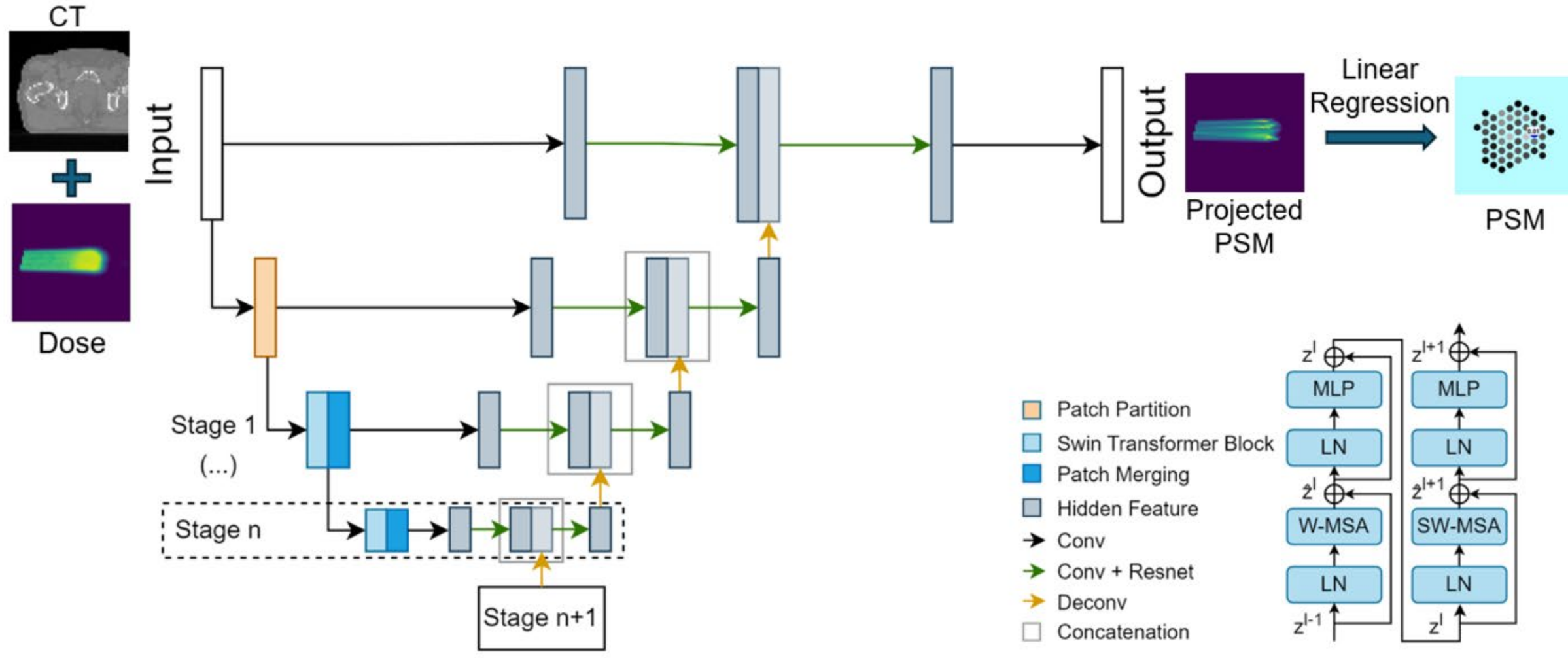

Figure 3. Architecture of the SwinUNETR model. The Swin transformer block is shown at bottom right. MLP: multilayer perceptron, LN: layer normalization, W-MSA and SW-MSA: window-based and shifted-window multihead self-attention.

2.3.2 Training Strategy

All model development and training were implemented in PyTorch and executed on four NVIDIA A100 GPUs with fixed random seeds. The cohort was split into training, validation, and test sets in an 80%/10%/10% ratio, with all fields from a given plan assigned to the same subset to avoid data leakage. CT, dose, and PrPSM volumes in all subsets underwent the same preprocessing described in Section 2.2.

During training, we applied spatial augmentations jointly to the CT/dose inputs and their PrPSM targets. Each sample had a 50% probability of a flip along the z axis and an 80% probability of a 3D affine transformation allowing translations of up to 10 voxels in the first two axes and 2 voxels in the third, and rotations of up to 45° about the z axis. Flips and rotations about the x and y axes were excluded because they would disrupt beam-path alignment with z and alter the WET characteristics that define the PrPSM.

The training objective was to minimize an L1 (absolute error) loss with dose-informed regional weighting. For each voxel, we computed the absolute difference between predicted and ground-truth PrPSM and then applied a binary mask derived from the normalized dose: voxels with normalized dose > 0.001 were assigned mask value 1, others 0. The final loss was the weighted sum of the mean absolute error (MAE) in the nonzero-dose region and the MAE in the zero-dose region, thus deprioritizing voxels that receive no dose and emphasizing clinically relevant regions. The same loss on the validation set was used for monitoring and early stopping:

$$\mathcal{L} = \frac{1}{|\mathcal{R}_+|}\sum_{i\in\mathcal{R}_+}|\hat{y}_i - y_i| + b\,\frac{1}{|\mathcal{R}_0|}\sum_{i\in\mathcal{R}_0}|\hat{y}_i - y_i|, \tag{2}$$

where $\mathcal{L}$ denotes the dose-informed loss function, $\hat{y}_i$ is the predicted PrPSM at voxel $i$, $y_i$ is the ground-truth PrPSM at voxel $i$, $\mathcal{R}_+$ is the set of voxels with normalized dose greater than 0.001 (nonzero-dose region), $\mathcal{R}_0$ is the set of voxels with normalized dose less than or equal to 0.001 (zero-dose region), $|\mathcal{R}_+|$and $|\mathcal{R}_0|$ denote the number of voxels in the nonzero-dose and zero-dose regions respectively, and the factor $b$ is a hyperparameter that down-weights the contribution of the loss in the zero-dose region to emphasize clinically relevant voxels receiving non-negligible dose.

Optimization used Adaptive Moment Estimation (Adam) with an initial learning rate of 0.002, a momentum parameter of 0.9, and cosine annealing. Models were trained with a batch size of 8 for up to 3000 epochs, with early stopping if the validation loss did not improve for 300 consecutive epochs. Regularization included dropout of 0.2 and weight decay of $1 \times 10^{-5}$. Hyperparameters such as learning rate, regional weighting factor, dropout, and weight decay were selected from a small grid of candidate values based on validation performance.

### 2.3.3 Ablation Analysis

We conducted an ablation study to assess the impact of the loss function and network architecture on PrPSM prediction performance.

For the loss function, we compared the proposed dose-informed, region-weighted L1 loss with two alternatives: a standard voxelwise L1 loss without dose-based masking or regional weighting, and a standard voxelwise L2 loss. For the architectural comparison, we evaluated SwinUNETR against two convolutional baselines: a 3D UNet with comparable depth and channel capacity,[21] and a 3D convolutional autoencoder without skip connections.[22] All models were trained using the same inputs, preprocessing, dataset split, and optimization settings to ensure a fair comparison.

## 2.4 Linear Regression-based Reconstruction of PSM from PrPSM

Under the straight-ray assumption and Equation (1), the PrPSM in each in-plane column can be written as a linear combination of the proton numbers in the contributing spots. This linearity enables approximate reconstruction of the original PSM from a PrPSM using linear regression.

Within GenSpot, this step converts the predicted PrPSM into a physically interpretable spot map; the resulting field-specific reconstruction is referred to as the GenSpot PSM.

As illustrated in Figure 4, reconstruction was performed independently for each in-plane column at location (x, y). For each selected column, we computed the WET along the beam direction on a voxel-by-voxel basis, forming a one-dimensional WET array. PDD curves for all clinically commissioned energies were stored as lookup tables and stacked to form the design matrix for regression.

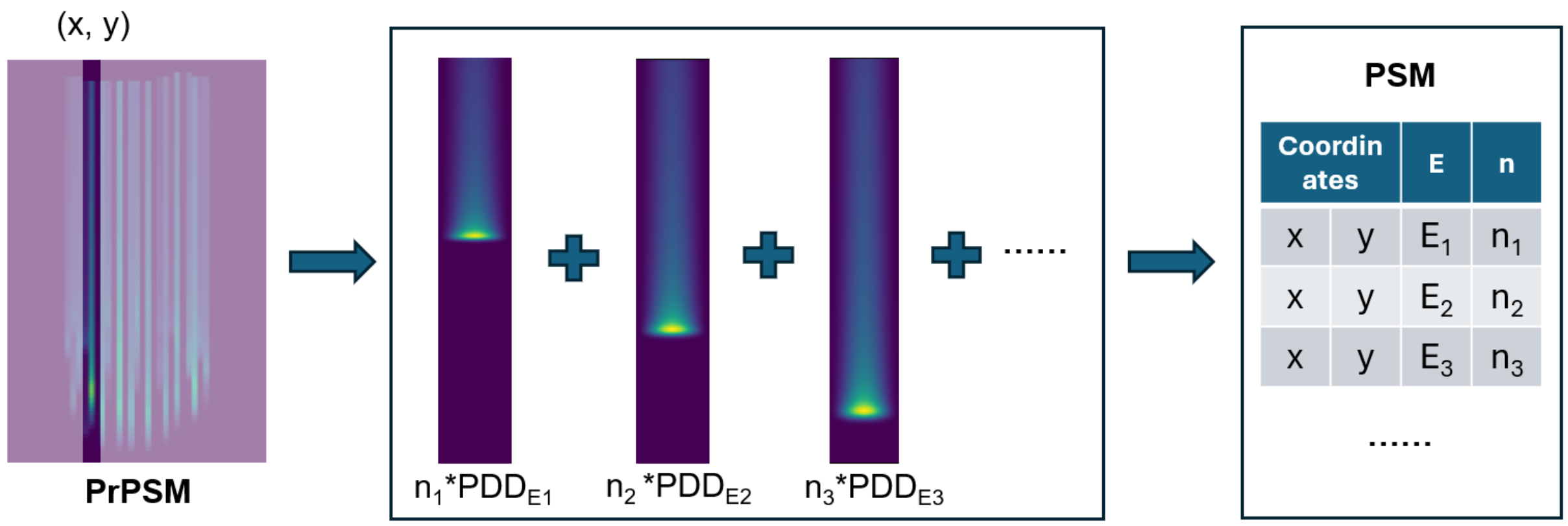

Figure 4. PSM reconstruction from PrPSM using linear regression.

Instead of ordinary least-squares regression, we used Lasso regression with non-negativity constraints to encourage sparse and physically plausible solutions.[23] This was implemented in Python using the LassoCV package from scikit-learn (version 1.3.0), with positivity enforced. The regularization parameter α was selected via five-fold cross-validation over a predefined range, performed independently for each column by partitioning the voxels in the selected WET interval into five contiguous depth segments. After fitting, coefficients corresponding to spot weights below the machine minimum (equivalent to 0.001 MU) were set to zero to satisfy delivery constraints and avoid unrealistically small spot contributions.

For each field, applying the SwinUNETR-based PrPSM prediction (Section 2.3) followed by Lasso reconstruction yields one GenSpot PSM that is compatible with the clinical delivery system.

**2.5 Evaluation and Statistical Analysis**

The performance of the GenSpot workflow was evaluated at two stages using the held-out test dataset: (i) PrPSM prediction accuracy and (ii) dose reproduction from reconstructed PSMs. All evaluations were performed under nominal beam parameters identical to those of the original clinical plans; robustness against setup or range uncertainties was not assessed.

First, we quantified how closely the PrPSM predicted from CT and dose matched the reference PrPSM obtained by projecting the clinically delivered PSM. Primary metrics included voxel-wise mean absolute error (MAE) within the non-zero-dose region, the structural similarity index measure (SSIM) and normalized cross-correlation (NCC). SSIM, which captures similarity in luminance, contrast and structural patterns, was computed for each two-dimensional slice along the third dimension and averaged across slices. NCC quantified the linear correlation between the predicted and reference PrPSM volumes. Because PrPSMs consist of straight, discrete, beam-aligned columns rather than physical dose distributions, SSIM and NCC were considered more informative than gamma analysis for this intermediate representation.

The second evaluation focused on the dose distribution resulting from the reconstructed PSM. For each field, the GenSpot PSM was transported through the clinical MC simulator to compute the corresponding dose distribution. Because all learning tasks used normalized PrPSM and PSM values, reconstructed spot weights were rescaled using the original field maximum dose before MC calculation so that absolute dose comparisons were meaningful. We compared the dose from the GenSpot PSM with the dose from the clinical PSM at both the individual-field and composite-plan levels, where field doses were summed to form the plan dose distribution. Evaluation metrics included voxel-wise MAE, 3D global gamma analysis (3% dose difference, 3 mm distance to agreement, 10% low-dose threshold) and dose-volume histogram (DVH) metrics for the CTV, bladder, and rectum. We did not directly compare the reconstructed PSM with the clinical PSM because the inverse mapping from dose to spot configuration is nonunique; our objective was dose reproduction rather than exact recovery of the clinical spot pattern.

## 3 Results

### 3.1 Accuracy of PrPSM prediction from CT and Dose

As described in Section 2.3.3, we compared multiple network architectures and loss functions for PrPSM prediction. Table 1 summarizes performance for SwinUNETR, a single-stage 3D UNet, and then 3D autoencoder under different voxel-wise loss functions.

| **Model** | **SwinUNETR** | | | **Single-stage UNet** | **Autoencoder** |
|---|---|---|---|---|---|
| **Loss Function** | Masked L1 | L1 | L2 | L1 | L1 |
| **MAE** | 0.06 ± 0.02 | 0.10 ± 0.02 | 0.12 ± 0.02 | 0.19 ± 0.05 | 0.24 ± 0.07 |
| **SSIM** | 0.94 ± 0.03 | 0.92 ± 0.06 | 0.92 ± 0.03 | 0.85 ± 0.06 | 0.79 ± 0.05 |
| **NCC** | 0.86 ± 0.10 | 0.82 ± 0.13 | 0.81 ± 0.09 | 0.77 ± 0.05 | 0.74 ± 0.04 |

Table 1. Performance of the evaluated models for PrPSM prediction.

The SwinUNETR trained with the proposed masked L1 loss achieved the best overall agreement with the reference PrPSM, with mean MAE 0.06 ± 0.02 (normalized units), SSIM 0.94 ± 0.03, and mean NCC 0.86 ± 0.10. The single-stage UNet trained with standard L1 loss performed slightly worse, with modest degradation across all three metrics. Autoencoder-based models showed substantially inferior performance, indicating reduced fidelity in reproducing the projected PSMs. These results suggest that both the transformer-based encoder-decoder and the dose-informed regional weighting in the loss function contribute meaningfully to PrPSM prediction accuracy. On the basis of this ablation analysis, the SwinUNETR trained with masked L1 loss was selected for the final GenSpot pipeline.

### 3.2 Performance of PSM Reconstruction and Dose Reproduction

For each field in the test set, the predicted PrPSM was converted to a GenSpot PSM using Lasso regression, and both the GenSpot PSM and the clinical PSM were transported through the MC simulator. Figure 5 illustrates a representative example. At the field level, the dose from the clinical PSM (Fig. 5a) and that from the GenSpot PSM (Fig. 5b) show similar entrance plateaus and distal fall-off, and the corresponding one-dimensional profiles along the dashed line nearly overlap (Fig. 5c). At the composite plan level, the clinical dose (Fig. 5d) and the dose from GenSpot PSM (Fig. 5e) are again visually indistinguishable, with closely matched profiles across the entrance, high-dose, and exit regions (Fig. 5f).

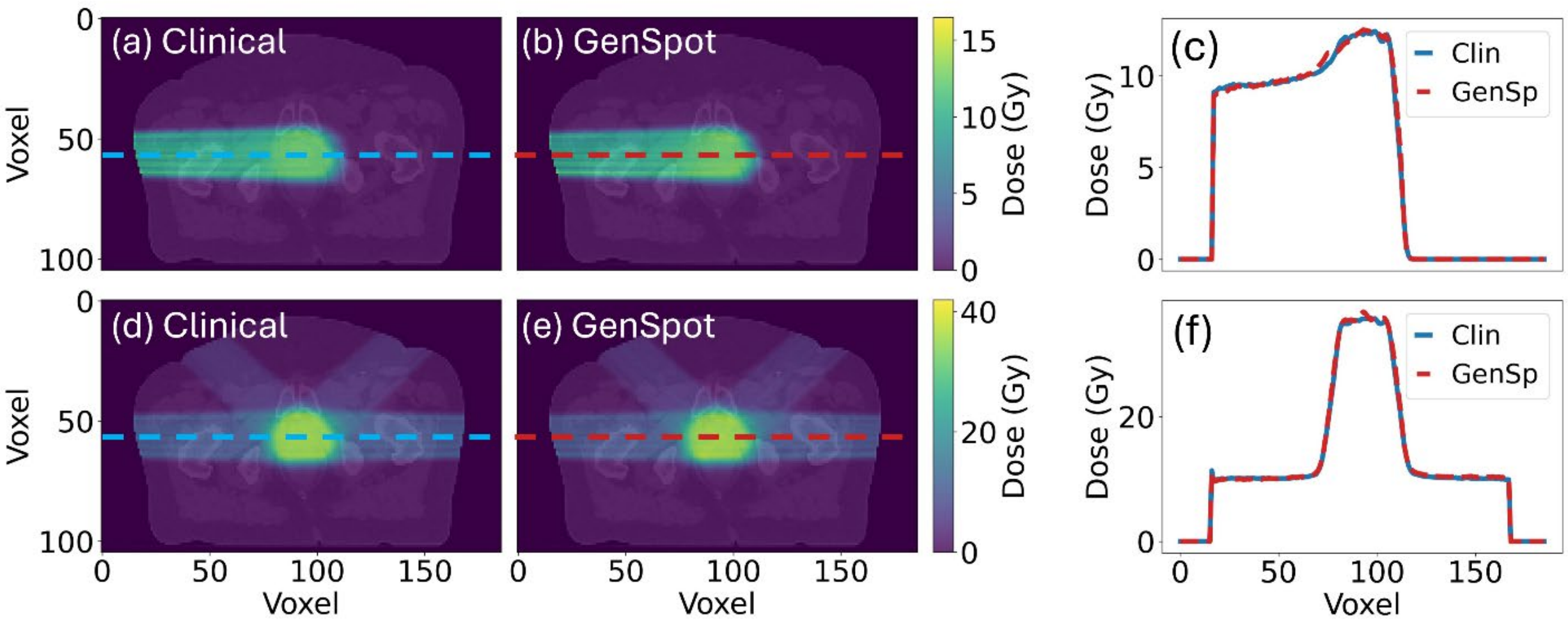

Figure 5. Representative comparison between the dose from clinical PSM (Clin) and the dose from GenSpot PSM (GenSp). (a) Field-level reference dose and (b) GenSpot dose, with the dashed line indicating the profile location; (c) corresponding dose profiles. (d) Composite plan reference dose and (e) composite dose from GenSpot PSM; (f) corresponding profiles.

Across the test dataset, field-level dose reproduction was accurate. Within the CTV, the dose from GenSpot PSM had an MAE of 0.30 ± 0.10 Gy relative to the dose from clinical PSM. Across the entire non-zero dose region, the MAE was 0.07 ± 0.03 Gy and the mean gamma passing rate (3%/3 mm) was 0.90 ± 0.05. At the composite-plan level, the CTV MAE was 0.82 ± 0.16 Gy, while the MAE within the non-zero dose region remained 0.07 ± 0.03 Gy. The corresponding plan-level gamma passing rate was 0.97 ± 0.03, indicating excellent agreement between the dose from the GenSpot PSM and the dose from the clinical PSM.

To quantify clinically relevant dosimetric differences, we computed DVH metrics for all test cases (Table 2). For the CTV, we evaluated $D_{mean}$, $D_{95\%}$, and $D_{5\%}$; for the rectum and bladder, we evaluated $D_{mean}$ and $D_{5\%}$. Table 2 reports the values for doses from the clinical and GenSpot PSMs, together with their per-patient difference (GenSpot minus clinical).

| Structure | Metric | Clinical PSM dose (Gy) | GenSpot PSM dose (Gy) | Δ dose (GenSpot − clinical) (Gy) |
|---|---|---|---|---|
| CTV | $D_{mean}$ | 38.56 ± 1.05 | 39.00 ± 1.05 | 0.43 ± 0.18 |
| | $D_{95\%}$ | 37.67 ± 1.00 | 37.41 ± 1.00 | -0.27 ± 0.36 |
| | $D_{5\%}$ | 39.31 ± 1.14 | 40.58 ± 1.14 | 1.26 ± 0.28 |
| Rectum | $D_{mean}$ | 4.78 ± 2.06 | 4.79 ± 2.12 | 0.01 ± 0.46 |
| | $D_{5\%}$ | 22.35 ± 7.61 | 21.75 ± 7.25 | -0.60 ± 1.22 |
| Bladder | $D_{mean}$ | 8.64 ± 3.47 | 8.39 ± 3.42 | -0.25 ± 0.73 |
| | $D_{5\%}$ | 35.43 ± 4.12 | 34.52 ± 3.93 | -0.91 ± 0.83 |

Table 2. DVH comparison between doses from the clinical and GenSpot PSMs.

Figure 6 shows representative DVHs for the CTV, bladder, and rectum. The CTV curves (Fig. 6a) nearly overlap, with a small rightward shift in the high-dose tail consistent with the positive $\Delta D_{5\%}$ reported in Table 2. For the bladder (Fig. 6b), the curves coincide across nearly the full dose range. For the rectum (Fig. 6c), the GenSpot curve is slightly lower in the low-dose region and converges with the clinical curve at higher dose. Overall, the DVHs demonstrate close dosimetric agreement across all three structures, although the systematic increase in CTV $D_{5\%}$ suggests a small but consistent hot-spot tendency that warrants consideration in SBRT protocols and should be assessed in the context of institutional dose constraints.

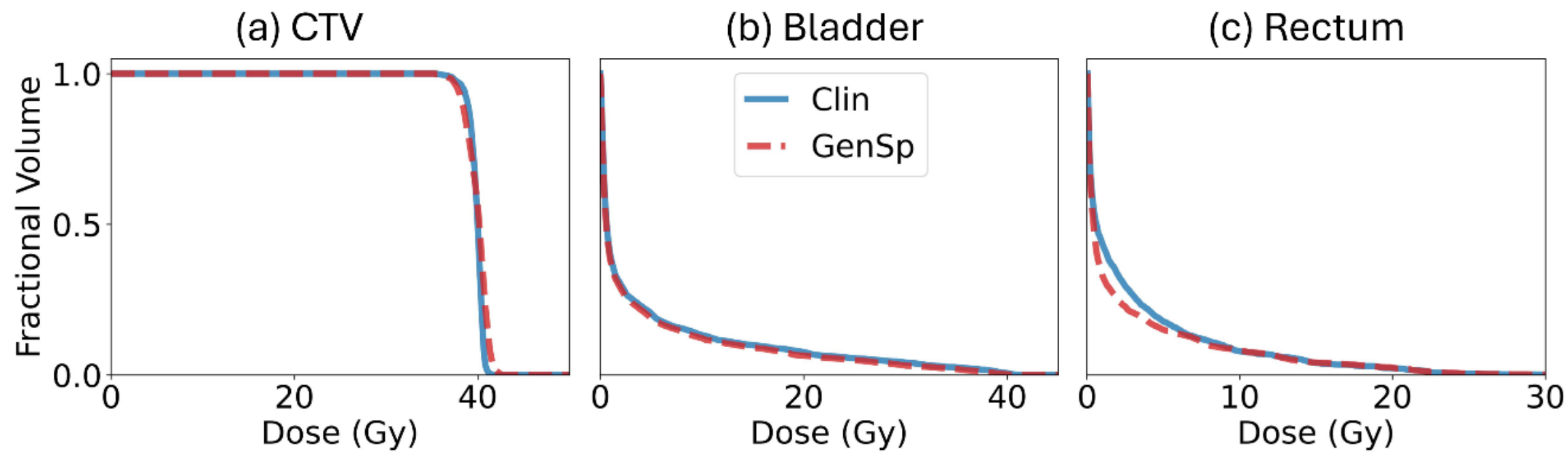

Figure 6. DVH comparison between dose from the clinical PSM (Clin) and dose from GenSpot PSM (GenSp) for (a) CTV, (b) bladder, and (c) rectum.

In addition to dosimetric agreement, we evaluated deliverability-related complexity metrics of the reconstructed spot maps. We compared the distributions of the number of nonzero spots and the number of nonzero energy layers between clinical and GenSpot PSMs. Across the test set, GenSpot reproduced complexity patterns similar to those of the clinical plans. For more than 90% of the reconstructed fields, the number of energy layers per field was within ±5 layers of the corresponding clinical field. GenSpot tended to produce a slightly higher number of non-zero spots than the clinical plans, with average spot counts of approximately 6000 versus 5000 spots. All reconstructed spot positions and energies fell within machine delivery limits by construction. A more detailed analysis of delivery time and log-file consistency was beyond the scope of this study and will require future work.

The computational efficiency of the GenSpot pipeline was also evaluated. On the A100 GPU used in this study, PrPSM prediction required approximately 0.02 s per field. Reconstruction of the GenSpot PSM from the predicted PrPSM via Lasso regression required about 2.1 s per field on CPU. The subsequent MC dose calculation remained the dominant computational cost, requiring approximately 4 min per field with the clinically commissioned GPU-based dose engine. Thus, the learned components of GenSpot contribute only a small fraction of total runtime, and further reductions in end-to-end latency will depend primarily on the choice and acceleration of dose calculation.

## 4 Discussion

In this study we introduced GenSpot, a two-stage framework that reconstructs deliverable proton spot maps (PSMs) directly from CT and three-dimensional dose for prostate SBRT plans treated with PBS. GenSpot combines a physics-informed projected proton spot map (PrPSM) representation with a transformer-based network for PrPSM prediction and a sparse linear reconstruction step for spot recovery. In a single-institution cohort of 259 prostate SBRT plans, MC doses computed from GenSpot PSMs closely match those computed from the clinically delivered PSMs, with low voxel-wise dose errors, high gamma passing rates, and small DVH deviations for the CTV, rectum, and bladder. Of note, GenSpot is not intended as a universal solver; rather, it is an inverse workflow designed to infer PSMs from clinically relevant dose distributions for a specific disease site.

At the PrPSM prediction stage, SwinUNETR outperformed both the UNet and autoencoder baselines, highlighting the importance of global contextual modeling and dose-informed loss weighting. The PrPSM representation itself is central to the approach: it embeds basic beam physics through WET and PDD while aligning naturally with voxel-based CT and dose grids, yet remains linearly related to the underlying spot intensities. This makes PrPSM a practical target for image-based learning and an effective bridge to subsequent linear reconstruction.

At the dose level, the reconstructed GenSpot PSMs reproduced the clinical MC dose with high fidelity. For composite plans, the mean absolute error in the non-zero dose region remained approximately 0.07 Gy, and the mean 3D gamma passing rate (3%/3 mm, 10% threshold) reached 0.97. DVH analysis showed preserved target coverage, with a modest increase in CTV $D_{5\%}$ of 1.3 Gy and a slight reduction in $D_{95\%}$. For rectum and bladder, mean and high-dose metrics differed by less than 1 Gy on average. This pattern suggests a small tendency toward hotter high-dose regions in the target while maintaining overall coverage and OAR sparing. Depending on institutional SBRT constraints, these deviations may be acceptable or may motivate additional postprocessing or local refinement.

The computational profile is also favorable. PrPSM prediction added 0.02 s per field, and regression-based reconstruction 2.1 s per field on the tested hardware, compared with roughly 4 min per field for MC dose computation. This suggests that, if combined with a faster dose engine or approximate dose calculation, GenSpot itself is unlikely to be the bottleneck in a time-sensitive workflow. In the current implementation, dose calculation dominates the total runtime regardless of whether the candidate PSM is generated by GenSpot or by conventional optimization.

GenSpot builds on our prior work showing that CT, PrPSM, and dose are strongly coupled and that a transformer can accurately learn the forward mapping from PrPSM and CT to dose. [17] The present study addresses the inverse problem by learning PrPSM from CT and dose and then recovering a deliverable PSM through sparse linear regression. Because the inverse problem is substantially more ill-posed than the forward problem, this is a critical step toward establishing a practical dose-to-spots link between AI dose-prediction models and machine-deliverable PBS treatment plans. In principle, GenSpot could be used to generate an initial PSM from an acceptable dose that can then be refined by conventional optimization, or to reconstruct updated spot maps from adapted dose distributions in online or offline adaptive workflows. In this study, however, the input dose to GenSpot was always the clinical MC dose, and the output was not treated as a clinically autonomous plan. The goal was to isolate and evaluate the dose-to-spots mapping under idealized conditions in which the target dose was already known to be acceptable.

This study has several limitations. First, it evaluated prostate SBRT plans from a single institution and a single PBS platform; the results should therefore be interpreted as a site- and machine-specific feasibility demonstration rather than proof of broad generalizability. Second, the PrPSM representation uses a straight-ray, WET-based approximation that neglects lateral scattering and range uncertainty, which appears adequate in the relatively homogeneous pelvis but may be insufficient in more heterogeneous anatomies. Third, all evaluations were performed under nominal conditions using the same beam model for both clinical and GenSpot PSMs, and we did not assess robustness to setup variation, anatomical change, or stopping-power uncertainty. Fourth, we did not compare GenSpot directly against dose-mimicking optimization or other dose-to-spots baselines, nor did we comprehensively evaluate delivery efficiency beyond simple measures such as spot count and number of energy layers. Finally, our analysis considered conventional physical dose and did not assess biological dose, which may differ between spot configurations that produce similar physical dose but different linear energy transfer patterns.[24]

These limitations point to several directions for future work. Evaluating GenSpot in additional anatomic sites and across different proton delivery systems will be necessary to determine where the current PrPSM formulation is adequate and where more advanced physical models are needed. Integrating GenSpot with contemporary field-based proton dose prediction models[25] would enable end-to-end assessment of an anatomy-to-dose-to-spots pipeline and would allow errors in the upstream dose prediction to propagate through GenSpot to the final PSM and dose. Further study should also include robustness analysis, delivery efficiency, and patient-specific QA, for example through robust DVH evaluation under setup and range uncertainties, treatment-time estimation, and phantom- or log-file-based verification of GenSpot-derived plans. In addition, methodological refinements such as alternative regularization strategies to better control spot-map complexity may further improve performance and scalability.

Taken together, these results show that, in this prostate SBRT cohort, GenSpot can infer a deliverable spot map from CT and dose such that the resulting MC dose closely matches that of the clinically delivered plan. While this work does not yet establish a fully general dose-to-spots solution, it demonstrates a practical and computationally efficient bridge between dose-based AI models and machine-deliverable PBS spot patterns that can be further refined and validated for broader clinical use.

## 5 Conclusion

GenSpot is a two-stage dose-to-spots framework that predicts a physics-informed projected proton spot map (PrPSM) from CT and dose using SwinUNETR and then reconstructs a machine-deliverable PBS spot map via sparse, non-negative Lasso regression. In 259 prostate SBRT plans, MC doses computed from GenSpot-reconstructed spot maps closely matched those from the clinically delivered spot maps, with low dose errors, high field- and plan-level gamma passing rates, and fast inference and reconstruction (0.02 s and 2.1 s per field, respectively). These findings support the use of GenSpot as an efficient bridge between dose-based models and deliverable PBS spot patterns, while further validation is needed across disease sites, upstream dose inputs, delivery constraints, and baseline dose-to-delivery approaches.